\documentclass[aip,jcp,reprint,amsmath,amssymb]{revtex4-2}

\usepackage[T1]{fontenc}
\usepackage[utf8]{inputenc}
\usepackage{lmodern}
\usepackage{microtype}
\usepackage{amsmath,amssymb,mathtools}
\usepackage{bm}
\usepackage{booktabs}
\usepackage[hidelinks]{hyperref}

\newcommand{\dd}{\mathrm d}
\newcommand{\E}{\mathbb E}

\newcommand{\diag}{\operatorname{diag}}
\newcommand{\cme}{\mathcal L}
\newcommand{\pathlaw}{\mathbb P}
\newcommand{\trans}{^{\mathsf T}}

\begin{document}

\title{Limits of Inferring Parametric Response from Single-Condition Trajectories in Stochastic Reaction Networks}

\author{Quentin Thommen}
\email{quentin.thommen@univ-lille.fr}
\thanks{ORCID: \href{https://orcid.org/0000-0002-0373-9199}{0000-0002-0373-9199}}
\affiliation{CRCLille--Cancer Research Center of Lille, Universit\'e de Lille, UMR 9020 CNRS, Inserm U1366, CHU de Lille, France}

\date{\today}

\begin{abstract}
Even exact knowledge of the Markov generator at one environmental condition does not determine how that generator changes with the environment unless additional reaction-structure information is supplied. We establish this non-identifiability for mass-action reaction networks driven by a chemostatted species. Distinct chemical embeddings can share the same generator and path law at a reference concentration \(u_0\) while differing in how the chemostat enters individual reaction channels. Changing integer chemostat reactant molecularities and rescaling the corresponding kinetic constants preserves all propensities at \(u_0\). This ambiguity already occurs for low-molecularity elementary reactions: a one-species birth--death example has identical path measures at the reference condition but dimensionless stationary sensitivities one and zero. Single-condition data can therefore determine the observed Markov dynamics at \(u_0\) without determining their parametric response. As an illustration of the complementary role of prior chemical structure, we also show that a known stoichiometric integral coordinate imposes a conditional zero-frequency identity within the linear-noise approximation.
\end{abstract}

\maketitle

\section{Introduction}
\label{sec:introduction}

Stochastic reaction networks are widely used to infer biochemical kinetics from molecular fluctuations and trajectory-resolved measurements. The chemical master equation (CME) provides the standard probabilistic description \cite{Gillespie1977}, while modern inference methods estimate kinetic parameters from stationary or time-resolved data \cite{MunskyKhammash2006,GuptaMikelsonKhammash2017,KomorowskiCostaRandStumpf2011,LiBarahonaThomas2025}. A complementary literature computes parameter sensitivities once a parametric reaction model \(a_r(x;\theta)\) has been specified \cite{Rathinam2010,Sheppard2012,Srivastava2013,Durrenberger2019,PantazisKatsoulakis2013,Ruess2017}.

These tasks require different information. Parameter inference concerns the generator of the observed Markov process at a measured condition, whereas sensitivity analysis concerns the derivative of a specified family of generators. The inverse problem considered here asks whether knowledge of
\begin{equation}
\cme_{u_0}
\qquad\text{determines}\qquad
\left.\partial_u\cme_u\right|_{u_0}.
\label{eq:intro-question}
\end{equation}
We assume that the Markov generator at \(u_0\) is known exactly, but that the reaction-channel incidence of the external species \(U\), including its reactant molecularity in individual elementary steps, is not independently specified. Equality of generators need not imply a unique decomposition into reaction channels; throughout, identifiability of \(\cme_{u_0}\) refers to the observed Markov dynamics, not to a unique chemical realization.

This distinction matters for stationary environmental response. Perfect stationary adaptation corresponds to vanishing stationary sensitivity and is associated with several biochemical control architectures, including integral feedback \cite{Yi2000,BriatGuptaKhammash2016,Araujo2023}. Recent fluctuation-response and trajectory-score approaches can compute responses from spontaneous trajectories when the perturbation direction is specified \cite{ZhengLu2025,Ptaszynski2026,Kwon2025,Aslyamov2026}. We ask the preceding inverse question: can the perturbation direction itself be inferred from data confined to \(u_0\)?

We show that the answer is negative in general over the stated mass-action class. The key point is that the ambiguity survives mass-action constraints and already occurs for low-molecularity elementary reactions: two chemically realizable networks can define exactly the same observed stochastic process at \(u_0\) while assigning different environmental sensitivities to the same observable. Changing the chemostat reactant molecularity of a reaction channel and rescaling its kinetic constant can preserve every propensity at \(u_0\). We call each chemically admissible choice of how \(U\) enters the reaction channels an environmental embedding of the internal dynamics. The resulting models are two chemically distinct environmental embeddings of the same stochastic dynamics at the reference condition. They satisfy
\begin{equation}
\widetilde{\cme}_{u_0}=\cme_{u_0},
\qquad
\widetilde{\pathlaw}_{u_0}=\pathlaw_{u_0},
\end{equation}
while generally
\begin{equation}
\left.
\partial_u\widetilde{\cme}_u
\right|_{u_0}
\neq
\left.
\partial_u\cme_u
\right|_{u_0}.
\end{equation}
A one-species birth--death example makes the consequence explicit: the two processes are statistically indistinguishable at \(u_0\), yet their dimensionless stationary sensitivities are one and zero.

This ambiguity is distinct from partial observation. Coarse graining and hidden states can make \(\cme_{u_0}\) difficult to identify \cite{WuJia2025,Goerlich2026,Marmolejo2026}; our construction persists even when that difficulty is removed by assuming \(\cme_{u_0}\) known exactly. The main result is therefore a non-injectivity of the map from admissible local families \(\{\cme_u\}_{u\approx u_0}\) to their value \(\cme_{u_0}\).

A short structural example at the end illustrates the complementary role of prior chemical information: for a network with a known stoichiometric integral coordinate, the LNA yields a conditional zero-frequency identity.

The paper is organized as follows. Section~\ref{sec:framework} states exactly what is learned from a trajectory at one condition. Section~\ref{sec:nonident} gives the minimal mass-action construction, and Sec.~\ref{sec:birthdeath} reduces it to an exact birth--death example. Section~\ref{sec:information} discusses which additional information resolves the ambiguity. Section~\ref{sec:stoich} gives the structural LNA example, followed by the discussion and conclusions.

\section{What a trajectory at one condition determines}
\label{sec:framework}

Consider internal molecular species $X=(X_1,\ldots,X_n)$ coupled to an environmental species $U$ held at a chemostatted concentration $u>0$, as in standard descriptions of open reaction networks \cite{RaoEsposito2016}.  Reaction channel $r$ changes the internal state by $s_r\in\mathbb Z^n$ and, under elementary mass action, has propensity
\begin{equation}
 a_r(x;u)=k_r u^{m_r} h_r(x),
 \label{eq:massaction}
\end{equation}
where $m_r\in\mathbb N_0$ is the reactant molecularity of the chemostatted species in that step and $h_r(x)$ contains the factors involving the internal species. This factorization is the standard mass-action separation of the chemostatted reactant factor from the internal-state dependence.  The CME generator is
\begin{equation}
 (\cme_u f)(x)=\sum_{r=1}^R a_r(x;u)\,[f(x+s_r)-f(x)].
 \label{eq:generator}
\end{equation}
For fixed $u$, this operator determines the continuous-time Markov process of the internal molecular counts and hence the probability law of every trajectory observable. In particular, if two models have the same generator at $u_0$, no statistic computed from trajectories recorded only at $u_0$ can distinguish them.

\begin{table}[t]
\caption{\label{tab:notation}Main notation used in the text.}
\centering
\begin{tabular}{@{}p{0.22\columnwidth}p{0.68\columnwidth}@{}}
\toprule
Symbol & Meaning \\
\midrule
$u,\;u_0$ & Environmental concentration and reference value \\
$X,\;x$ & Internal molecular-count process and one of its states \\
$s_r$ & Stoichiometric jump of reaction channel $r$ \\
$a_r(x;u)$ & Propensity of reaction channel $r$ \\
$k_r$ & Kinetic constant of reaction channel $r$ \\
$m_r$ & Chemostat reactant molecularity in channel $r$ \\
$\cme_u$ & Generator of the observed Markov process at concentration $u$ \\
$\pathlaw_u$ & Path law generated by $\cme_u$ \\
  $\chi_y$ & Dimensionless stationary sensitivity, $\left.\dd\ln\bar y/\dd\ln u\right|_{u_0}$\\
$J,\;D$ & LNA drift Jacobian and chemical diffusion matrix \\
$Z_y(\omega)$ & Bilateral power spectral density of the LNA output fluctuation \\
\bottomrule
\end{tabular}
\end{table}

To isolate the issue of interest from finite sampling and hidden-state effects, we take the ideal limit in which an experiment identifies $\cme_{u_0}$ exactly.  The remaining question is then not how well the dynamics at $u_0$ are known, but whether those dynamics determine how the environment enters the reactions.

They do not. A trajectory at $u_0$ fixes the effective state-to-state transition rates encoded in $\cme_{u_0}$, but not the power of $u$ associated with each admissible chemical embedding.  For Eq.~\eqref{eq:massaction},
\begin{equation}
 \left.\partial_{\ln u} a_r(x;u)\right|_{u_0}
 =m_r a_r(x;u_0).
 \label{eq:propensity-derivative}
\end{equation}
Thus the response to changing $u$ requires chemical information absent from the generator at a single condition: the chemostat incidence \(m_r\) of the reaction channels. For a positive stationary observable with mean $\bar y(u)$, we quantify the local stationary response by
\begin{equation}
\chi_y
=
\left.
\frac{\dd\ln \bar y(u)}{\dd\ln u}
\right|_{u_0}.
\label{eq:chi-def}
\end{equation}
The condition $\chi_y=0$ expresses local stationary insensitivity to \(u\); in adaptive systems it is the local signature of perfect stationary adaptation.

\section{Same dynamics at $u_0$, different dynamics away from $u_0$}
\label{sec:nonident}

The non-identifiability follows from a direct mass-action construction.  Start from Eq.~\eqref{eq:massaction} and choose another chemically admissible molecularity $\widetilde m_r$ for one or more channels.  Define the corresponding rate constant by
\begin{equation}
 \widetilde k_r=k_r u_0^{m_r-\widetilde m_r}.
 \label{eq:ktilde}
\end{equation}
The modified propensity is
\begin{equation}
 \widetilde a_r(x;u)=\widetilde k_r u^{\widetilde m_r}h_r(x).
\end{equation}
At the reference condition,
\begin{equation}
 \widetilde a_r(x;u_0)=a_r(x;u_0)
 \qquad\text{for all }x,r,
 \label{eq:equal-propensities}
\end{equation}
which is a sufficient channelwise condition for generator equality; the individual channel propensities need not themselves be observable. Thus the two networks have exactly the same generator,
\begin{equation}
 \widetilde\cme_{u_0}=\cme_{u_0}.
 \label{eq:equal-generators}
\end{equation}
Consequently they have the same path law at $u_0$, not merely the same mean, covariance, or spectrum.  However,
\begin{equation}
 \left.\partial_{\ln u}\widetilde a_r\right|_{u_0}
 =\widetilde m_r a_r(x;u_0),
 \qquad
 \left.\partial_{\ln u}a_r\right|_{u_0}
 =m_r a_r(x;u_0),
 \label{eq:different-derivatives}
\end{equation}
and therefore
\begin{align}
\left.
\partial_{\ln u}
(\widetilde{\cme}_u-\cme_u)f
\right|_{u_0}
&=
\sum_r
(\widetilde m_r-m_r)a_r(x;u_0)
\nonumber\\
&\quad\times
\bigl[f(x+s_r)-f(x)\bigr].
\label{eq:generator-derivative-difference}
\end{align}
The generator derivatives differ whenever the right-hand side does not cancel identically. Thus the same generator at $u_0$ can admit distinct chemically admissible environmental tangents. This is not an arbitrary smooth reparameterization freedom: the ambiguity is generated by a discrete change in integer chemostat reactant molecularity together with the rate-constant rescaling required by mass action.

\paragraph*{Proposition 1.}
\emph{Within the mass-action class of Eq.~\eqref{eq:massaction}, the
  trajectory law at a single condition $u_0$ does not in general
  identify the stationary response to $u$. If the chemostat reactant molecularities are not independently specified, two chemically distinct environmental embeddings can satisfy $\widetilde\cme_{u_0}=\cme_{u_0}$ while having different derivatives with respect to $u$. The birth--death construction below shows explicitly that their stationary responses can differ.}

\paragraph*{Proof.}
Equation~\eqref{eq:ktilde} gives Eq.~\eqref{eq:equal-propensities}, hence Eq.~\eqref{eq:equal-generators}. Equality of generators implies equality of the path law at $u_0$ for any common initial distribution. The two embeddings therefore have identical path measures at the reference condition and, in the language of stochastic trajectory distinguishability, zero path-space distinguishability \cite{Pagare2024}. Equation~\eqref{eq:generator-derivative-difference} shows that their environmental generator derivatives can differ. The birth--death example below gives an explicit observable whose stationary sensitivity differs, completing the construction. $\square$

\section{Exact birth--death example}
\label{sec:birthdeath}

The argument is already complete in a one-species network.  For this example, normalize the chemostat concentration by its reference value so that $u_0=1$.

In network A, $U$ affects production but not removal,
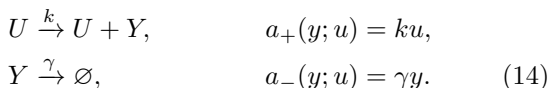
\begin{align}
U &\xrightarrow{k} U+Y, & a_+(y;u)&=ku,\nonumber\\
Y &\xrightarrow{\gamma}\varnothing, & a_-(y;u)&=\gamma y.
\label{eq:network-A}
\end{align}
The stationary mean is
\begin{equation}
 y_A^*(u)=\frac{k}{\gamma}u,
 \qquad
 \chi_A\equiv\frac{\dd\ln y_A^*}{\dd\ln u}=1.
 \label{eq:network-A-response}
\end{equation}

In network B, $U$ multiplies both production and removal,
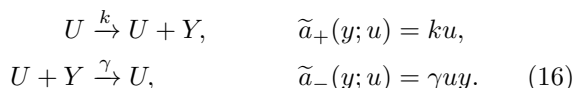
\begin{align}
U &\xrightarrow{k} U+Y, & \widetilde a_+(y;u)&=ku,\nonumber\\
U+Y &\xrightarrow{\gamma}U, & \widetilde a_-(y;u)&=\gamma uy.
\label{eq:network-B}
\end{align}
The factor $u$ rescales the entire birth--death generator, changing the dynamical timescale but not the stationary balance:
\begin{equation}
 y_B^*(u)=\frac{k}{\gamma},
 \qquad
 \chi_B=0.
 \label{eq:network-B-response}
\end{equation}
Here adaptation refers only to the stationary input--output property $\chi_B=0$, not to a feedback-control mechanism.

At $u_0=1$, however, both networks have exactly the same birth rate $k$ and death rate $\gamma y$.  Therefore
\begin{equation}
 \cme^{A}_{u_0}=\cme^{B}_{u_0},
\end{equation}
and every trajectory statistic is identical.  For example, both have the same stationary Poisson distribution,
\begin{equation}
 P(Y=n)=e^{-k/\gamma}\frac{(k/\gamma)^n}{n!},
 \label{eq:poisson}
\end{equation}
the same autocovariance,
\begin{equation}
 C_Y(t)=\frac{k}{\gamma}e^{-\gamma|t|},
 \label{eq:birthdeath-covariance}
\end{equation}
and the same bilateral power spectral density,
\begin{equation}
 S_Y(\omega)=\frac{2k}{\gamma^2+\omega^2}.
 \label{eq:birthdeath-psd}
\end{equation}
These equalities are illustrative: generator equality already guarantees equality of the full path law.

This example gives the physical content of Proposition~1.  A trajectory at $u_0$ learns the birth rate $k$, the death law $\gamma y$, and therefore the observed Markov dynamics at that condition.  It does not learn whether the observed death rate came from a channel proportional to $y$ or from a channel proportional to $uy$ evaluated at $u=1$.  The two networks are therefore exactly indistinguishable at $u_0$.  When $u$ changes, the hidden distinction becomes observable: network A changes its stationary output, whereas network B changes only its timescale.

\section{What additional information resolves the ambiguity?}
\label{sec:information}

The result separates two experimentally distinct tasks.  More data at fixed $u_0$ can improve the estimate of $\cme_{u_0}$; it cannot reveal how $\cme_u$ changes with $u$ when several chemical embeddings agree exactly at $u_0$.  The missing information must constrain the environmental direction itself.

Two direct routes constrain this missing direction.  First, independent chemical knowledge can specify which elementary reactions contain $U$ and with what molecularity.  Second, trajectories measured at two or more nearby values of $u$ directly probe how the transition rates change with the environment.  Either type of information can remove the particular ambiguity constructed above, subject to the usual identifiability conditions of the chosen model class.

This limitation is distinct from partial observation.  Coarse graining can make even the generator at $u_0$ difficult to recover \cite{WuJia2025}; hidden variables can induce memory \cite{Goerlich2026}; and model reduction can alter dynamical spectra \cite{Marmolejo2026}.  Here we remove that problem by granting exact knowledge of $\cme_{u_0}$.  The ambiguity remains because the missing object is not additional dynamics at $u_0$, but the chemical dependence of those dynamics on $u$.

The same point clarifies the relation to fluctuation--response formulas.  Trajectory-score methods can compute a response from unperturbed trajectories once the perturbation is specified \cite{ZhengLu2025}.  For a jump process, the score contains derivatives of transition rates with respect to the chosen parameter.  In Eq.~\eqref{eq:massaction}, those derivatives contain the molecularities $m_r$.  Such formulas solve the response problem once the environmental direction is specified; Proposition~1 asks whether that direction can itself be inferred from data restricted to one condition.

\section{Structural example: an integral coordinate constrains zero-frequency noise}
\label{sec:stoich}

The non-identifiability result concerns what cannot be inferred from trajectory data alone. We now consider the complementary situation in which independently known chemical structure constrains a fluctuation observable even though kinetic details orthogonal to the specified projection remain unknown. When an integral coordinate and its projected drift relation are known, chemical stoichiometry imposes a conditional identity on spontaneous fluctuations within the LNA.

\subsection{Linear-noise approximation for chemical reaction networks}
\label{subsec:lna}

We use the standard van Kampen normalization. For system size (or volume) \(\Omega\), the copy-number vector is written near a stable deterministic fixed point as
\begin{equation}
N(t)
=
\Omega\phi^*
+
\sqrt{\Omega}\,\xi(t),
\label{eq:vankampen-scaling}
\end{equation}
where \(\xi\) is the rescaled fluctuation variable. To leading order, the LNA gives the Ornstein--Uhlenbeck process \cite{ElfEhrenberg2003,Marmolejo2026}
\begin{equation}
\dot\xi
=
J\xi
+
S F^{1/2}\eta,
\qquad
F=\diag(v_1^*,\ldots,v_R^*),
\label{eq:lna}
\end{equation}
with
\begin{equation}
\left\langle
\eta(t)\eta\trans(t')
\right\rangle
=
\delta(t-t')I.
\end{equation}
Here \(v_r^*\) denotes the macroscopic stationary reaction flux in the same normalization, and the chemical diffusion matrix is
\begin{equation}
D=SFS\trans.
\label{eq:chemical-diffusion}
\end{equation}
Equation~\eqref{eq:chemical-diffusion}, which links reaction fluxes to fluctuation amplitudes, is the chemical-kinetic ingredient used below.

Throughout this section,
\begin{equation}
Z_\xi(\omega)
=
\int_{-\infty}^{\infty}
\left\langle
\xi(t)\xi(0)\trans
\right\rangle_{\rm ss}
e^{-i\omega t}\,\dd t
\label{eq:lna-psd-def}
\end{equation}
denotes the bilateral PSD of the rescaled LNA fluctuation. It is
\begin{equation}
Z_\xi(\omega)
=
(J-i\omega I)^{-1}
D
(J\trans+i\omega I)^{-1},
\label{eq:lna-spectrum}
\end{equation}
consistent with standard LNA spectral formulas \cite{ElfEhrenberg2003,Aslyamov2026}. For the rescaled output fluctuation
\(
y=c\trans\xi
\),
we write
\begin{equation}
Z_y(\omega)
=
c\trans Z_\xi(\omega)c.
\label{eq:output-spectrum}
\end{equation}
If the physical observable is the concentration fluctuation
\(
\delta\phi=N/\Omega-\phi^*=\xi/\sqrt{\Omega}
\),
its PSD is \(Z_y/\Omega\); the copy-number fluctuation
\(
\delta N=\sqrt{\Omega}\,\xi
\)
has PSD \(\Omega Z_y\). All zero-frequency identities below refer to \(Z_y\) in the rescaled convention of Eq.~\eqref{eq:lna-psd-def}.

\subsection{Integral stoichiometric coordinate}
\label{subsec:integral-coordinate}

Integral coordinates and more general embedded integral structures underlie broad classes of robustly adapting reaction networks \cite{Yi2000,BriatGuptaKhammash2016,Araujo2023}. Here we assume such a coordinate is independently known and ask what it constrains about the LNA spectrum.

Let \(q\trans X\) be a stoichiometric integral coordinate.
Assume that all internal reactions satisfy
\begin{equation}
q\trans s_r=0,
\label{eq:orthogonal-reactions}
\end{equation}
apart from two families of events.
Reference events change \(q\trans X\) by \(+1\) with total stationary flux \(j_+\), whereas sensing events change it by \(-1\) with total stationary flux \(j_-\).

At the deterministic level, suppose the projected linearized drift satisfies
\begin{equation}
q\trans J=-\theta c\trans,
\qquad
\theta>0.
\label{eq:qJ}
\end{equation}
The output direction is therefore linked to the time derivative of the integral coordinate. Stationarity of that coordinate imposes
\begin{equation}
j_+=j_-\equiv j.
\label{eq:flux-balance}
\end{equation}

At zero frequency, after restriction to independent coordinates when conservation laws are present,
\begin{equation}
Z_y(0)
=
c\trans J^{-1}DJ^{-\trans}c.
\label{eq:zero-output}
\end{equation}
Equation~\eqref{eq:qJ} gives
\begin{equation}
c\trans J^{-1}
=
-\frac{1}{\theta}q\trans,
\end{equation}
and hence
\begin{align}
Z_y(0)
&=
\frac{1}{\theta^2}
q\trans Dq
\nonumber\\
&=
\frac{1}{\theta^2}
\sum_r
v_r^*
(q\trans s_r)^2.
\label{eq:projection-noise}
\end{align}
Every reaction orthogonal to \(q\) drops out term by term. The two non-orthogonal event families contribute \(j_+\) and \(j_-\), so Eq.~\eqref{eq:flux-balance} yields
\begin{equation}
Z_y(0)
=
\frac{2j}{\theta^2}.
\label{eq:zero-invariant}
\end{equation}

Within the Ornstein--Uhlenbeck process defined by the LNA, Eq.~\eqref{eq:zero-invariant} follows directly from the stated structural assumptions. Additional reactions make no direct contribution to \(q\trans Dq\) when they are stoichiometrically orthogonal to \(q\). They may still modify the fixed point, Jacobian, and fluxes; the identity remains valid only when they also preserve the projected drift relation~\eqref{eq:qJ} and stability of the fixed point.

\subsection{Antithetic integral feedback}
\label{subsec:aif}

The antithetic integral feedback (AIF) controller introduced by Briat, Gupta, and Khammash provides a canonical stochastic biochemical realization of integral control \cite{BriatGuptaKhammash2016}. Consider
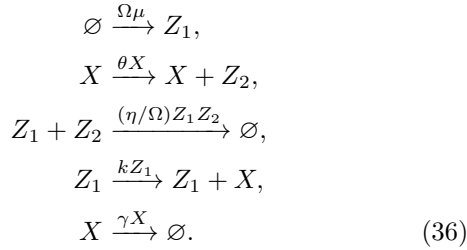
\begin{align}
\varnothing
&\xrightarrow{\Omega\mu}
Z_1,
\nonumber\\
X
&\xrightarrow{\theta X}
X+Z_2,
\nonumber\\
Z_1+Z_2
&\xrightarrow{(\eta/\Omega)Z_1Z_2}
\varnothing,
\nonumber\\
Z_1
&\xrightarrow{kZ_1}
Z_1+X,
\nonumber\\
X
&\xrightarrow{\gamma X}
\varnothing.
\label{eq:aif-reactions}
\end{align}
The coordinate
\begin{equation}
q\trans X_{\mathrm{full}}
=
Z_1-Z_2
\label{eq:aif-q}
\end{equation}
is unaffected by annihilation, actuation, and degradation of the regulated species. Its mean dynamics satisfy
\begin{equation}
\frac{\dd}{\dd t}
\E[Z_1-Z_2]
=
\Omega\mu-\theta\E[X].
\label{eq:aif-integral}
\end{equation}
Because this projected drift is already linear in \(X\), its linearization satisfies Eq.~\eqref{eq:qJ} exactly.
Provided a stationary regime with finite first moments exists,
\begin{equation}
\frac{\E[X]}{\Omega}
=
\frac{\mu}{\theta},
\label{eq:aif-setpoint}
\end{equation}
which is the characteristic adapted set point of the AIF architecture \cite{BriatGuptaKhammash2016}. This mean relation follows directly from the stochastic moment balance. The spectral result below additionally assumes that the associated deterministic fixed point is stable so that the LNA is well defined.

The set-point relation~\eqref{eq:aif-setpoint} is an exact stationary first-moment statement. The following spectral identity is separate: it is obtained after linearization about a stable deterministic fixed point. At the macroscopic level, the two events that change \(Z_1-Z_2\) carry equal stationary flux
\(
j=\mu
\),
and Eq.~\eqref{eq:zero-invariant} gives, for the rescaled LNA fluctuation of \(X\),
\begin{equation}
Z_{\xi_X}(0)
=
\frac{2\mu}{\theta^2}.
\label{eq:aif-zero}
\end{equation}

Noise and performance tradeoffs of AIF controllers have been studied extensively \cite{BriatGuptaKhammash2018,HilfingerNoise2023}, and frequency spectra have been used to characterize adaptive biochemical circuits \cite{FrequencySpectra2022}. Equation~\eqref{eq:aif-zero} concerns the zero-frequency spectral value that the stoichiometric integral coordinate fixes within the LNA.

The converse inference does not follow. The simple birth--death process
\begin{equation}
\varnothing
\xrightarrow{\Omega\mu}
Y,
\qquad
Y
\xrightarrow{\theta Y}
\varnothing
\label{eq:control-bd}
\end{equation}
has
\begin{equation}
\phi_Y^*=\frac{\mu}{\theta},
\qquad
Z_{\xi_Y}(0)=\frac{2\mu}{\theta^2}.
\label{eq:control-bd-result}
\end{equation}
Thus the pair \((\phi_Y^*,Z_{\xi_Y}(0))\) does not identify the AIF architecture. Equation~\eqref{eq:aif-zero} is therefore a prediction conditional on known integral structure, not a unique inverse signature.

\section{Discussion}
\label{sec:discussion}

The central distinction is between identifying the observed generator at \(u_0\) and identifying its environmental derivative. Standard sensitivity calculations supply the parametric family as part of the model and therefore do not face this inverse problem.

\subsection{Sensitivity analysis and chemical meaning}
\label{subsec:discussion-sensitivity}

Sensitivity analysis for stochastic reaction networks typically starts from a parameterized generator and asks how an observable changes when one of the specified parameters varies \cite{Rathinam2010,Sheppard2012,Srivastava2013,Durrenberger2019}. Path-space sensitivity methods encode the same specified perturbation through changes in trajectory measures, relative entropy rates, or Fisher information \cite{PantazisKatsoulakis2013,PantazisKatsoulakisVlachos2013}; methods designed for extrinsic variability likewise assume a defined dependence of propensities on environmental or population-level parameters \cite{Ruess2017}. In all of these settings, the perturbation direction is part of the model definition. Proposition~1 addresses a different inverse problem. The trajectory law at \(u_0\) may identify the stochastic dynamics at that condition, yet several chemically admissible environmental embeddings can realize those same dynamics at the reference condition while differing away from it.

The molecularity transformation isolates information that is invisible at \(u_0\): it preserves the stochastic process at the reference condition while changing its derivative with respect to \(u\). This differs from standard structural or practical identifiability of parameters within a fixed model: even an ideal procedure that perfectly identifies all kinetic quantities entering \(\cme_{u_0}\) still cannot determine the environmental continuation \(u\mapsto\cme_u\) without additional information. The result therefore sets a limit on what any single-condition inference procedure can recover without structural assumptions on the environmental channel \cite{KomorowskiCostaRandStumpf2011,LiBarahonaThomas2025}.

This distinction also clarifies response calculations from spontaneous trajectories. Score-based methods can evaluate sensitivities efficiently from unperturbed data \cite{ZhengLu2025}, but the score itself contains the derivative of the transition rates with respect to the chosen perturbation parameter. Our result concerns the identifiability of this derivative, rather than the subsequent computation of the response once it is given.

Mass-action kinetics restrict the environmental dependence of elementary propensities to chemically admissible molecularities. Proposition~1 shows that the usual interpolation ambiguity survives these restrictions. The birth--death example makes this concrete with two low-order environmental embeddings: one couples \(U\) only to production, the other to production and removal. The construction therefore compares distinct chemistries that realize the same observed Markov dynamics at the reference condition; it is not a reparameterization of one known reaction mechanism.

Coarse-grained observations create equivalence classes of microscopic generators consistent with the same observed trajectory statistics \cite{WuJia2025}. Model reduction can likewise preserve some low-order quantities while altering dynamical spectra and information transmission \cite{Marmolejo2026}. These are limitations on recovery of the dynamics at \(u_0\).

The molecularity ambiguity remains after that first problem is removed. In practical experiments, hidden states can make \(\cme_{u_0}\) uncertain while unknown environmental incidence can make \(\partial_u\cme_u|_{u_0}\) uncertain even for a fixed candidate generator.

\subsection{Structural information and experimental implications}
\label{subsec:discussion-structure}

The zero-frequency example illustrates the complementary role of specified structure: once a particular stoichiometric relation is known, spontaneous fluctuations can obey additional conditional identities. Mass-action chemistry links deterministic fluxes and stochastic shot noise through
\(
D=SFS\trans
\).
A stoichiometric coordinate can therefore remove whole classes of hidden reactions from a projected fluctuation observable. This mechanism is chemical rather than generic to arbitrary linear stochastic systems with independent drift and diffusion matrices.

The result is related in spirit to structural robustness in reaction networks \cite{ShinarFeinberg2010}, but the observable differs. Absolute concentration robustness concerns a stationary concentration value. Equation~\eqref{eq:zero-invariant} concerns the frequency-zero component of stochastic fluctuations around a stable state. The stochastic and deterministic notions should therefore be kept distinct, as emphasized more broadly by work on stochastic reaction networks with deterministic structural robustness \cite{AndersonEncisoJohnston2014}.

For experiments that seek to infer adaptation from spontaneous data, increasing trajectory length, temporal resolution, or the number of observed internal species improves knowledge of the dynamics at the measured condition but does not determine how an unobserved environmental variable enters the reaction channels. Parametric response becomes identifiable only when the experiment or prior chemistry constrains the relevant environmental direction.

This perspective also separates mechanistic and parametric-response inference. A structural model can predict a structural invariant without uniquely identifying all hidden kinetics, as in Eq.~\eqref{eq:zero-invariant}. Conversely, a single fluctuation invariant need not identify a unique molecular mechanism, as demonstrated by the birth--death control in Eqs.~\eqref{eq:control-bd}--\eqref{eq:control-bd-result}. Experimental claims should therefore state which direction of implication is supported by the available information.

\section{Conclusions}
\label{sec:conclusions}

Even exact knowledge of the Markov generator at one environmental condition does not, without additional reaction-structure information, determine its derivative with respect to that environment. Within the mass-action class considered here, this ambiguity persists under integer chemostat reactant molecularities and chemically consistent rate constants: distinct environmental embeddings can share the same generator and path law at \(u_0\) while producing different stationary sensitivities. The birth--death example realizes this difference with dimensionless sensitivities one and zero.

A response is identifiable from single-condition data only if all chemically admissible continuations consistent with those data give the same response. Measurements at multiple control values or independent knowledge of reaction incidence can reduce this ambiguity by constraining the environmental direction.

Known chemical structure can also impose predictive fluctuation constraints. For a stoichiometric integral coordinate satisfying the stated linearized relation, the LNA yields a zero-frequency identity determined by the projected reaction fluxes. This example complements the non-identifiability result by showing what becomes predictable once structural information beyond the single-condition trajectory is supplied.

The practical distinction is therefore simple: learning the Markov dynamics at one condition is not the same as learning how a physical control variable changes those dynamics. Sensitivity analysis begins once that environmental direction is specified or independently constrained.

\section*{Conflict of Interest}
The author has no conflicts to disclose.


\section*{Data Availability}
Data sharing is not applicable to this article as no new data were created or analyzed in this study.

\appendix

\section{Generator equality implies equality of all single-condition trajectory statistics}
\label{app:pathlaw}

For a continuous-time Markov jump process, the generator fixes the state-jump rates and escape rates. Hence two models with the same generator at \(u_0\) and the same initial distribution assign the same probability to every finite trajectory segment. Our construction imposes the stronger sufficient condition \(\widetilde a_r(x;u_0)=a_r(x;u_0)\) channel by channel, but the observational conclusion relies only on generator equality and does not assume that reaction-channel labels are observed. For an ergodic common generator, the stationary initial distribution is also the same. Consequently all single-condition trajectory functionals---including autocorrelations, power spectra, waiting-time distributions, occupation times, and jump counts---are identical.

\section{Zero-frequency invariant for general jump magnitudes}
\label{app:general-jumps}

The derivation in Sec.~\ref{sec:stoich} assumed unit changes of the integral coordinate for the reference and sensing events. Suppose instead that reaction family \(+\) changes \(q\trans X\) by \(+\nu_+\) and reaction family \(-\) by \(-\nu_-\), with positive integers \(\nu_\pm\). Stationarity gives
\begin{equation}
\nu_+j_+
=
\nu_-j_-.
\end{equation}
The projected chemical diffusion becomes
\begin{equation}
q\trans Dq
=
\nu_+^2j_+
+
\nu_-^2j_-.
\end{equation}
Using the stationarity relation,
\begin{equation}
q\trans Dq
=
\nu_+j_+
(\nu_++\nu_-)
=
\nu_-j_-
(\nu_++\nu_-).
\end{equation}
The output spectrum at zero frequency is therefore
\begin{equation}
Z_y(0)
=
\frac{
\nu_+^2j_+
+
\nu_-^2j_-
}{\theta^2}.
\end{equation}
Equation~\eqref{eq:zero-invariant} is recovered for
\(
\nu_+=\nu_-=1
\).

\section{Scope of the linear-noise result}
\label{app:lna-scope}

Equation~\eqref{eq:zero-invariant} uses the LNA around a stable fixed point in the rescaled convention of Eq.~\eqref{eq:vankampen-scaling}. Its ingredients are: (i) a chemical diffusion matrix of the form \(D=SFS\trans\), (ii) a stable and invertible Jacobian on the relevant independent coordinates, (iii) the projected relation \(q\trans J=-\theta c\trans\), and (iv) the specified stoichiometric action of the reference and sensing reactions on \(q\). In networks with conservation laws, the reduction to the stoichiometric compatibility class precedes the inversion of \(J\).

The result does not require complete knowledge of hidden kinetic details, but it does require the integral coordinate and its stoichiometric support to be known. Strongly non-Gaussian regimes, multimodal stationary distributions, and systems far from a stable fixed point fall outside the LNA derivation and should be treated separately.


\end{document}